\documentclass[12pt]{article}

\usepackage{amsmath}
\usepackage{amssymb}
\usepackage{cite}

\def\nn{\nonumber}

\numberwithin{equation}{section}

\title{\bf \Large Perfect fluid dynamics
with conformal\\ Newton-Hooke symmetries }

\author{Timofei  Snegirev${}^{a}$\thanks{timofei.v.snegirev@tusur.ru}
\\[0.5cm]
\it{\small ${}^a$Laboratory of Applied Mathematics and Theoretical Physics,}\\
\it{\small Tomsk State University of Control Systems and Radioelectronics,}\\
\it{\small Lenin ave. 40, 634050 Tomsk, Russia}}

\date{}

\begin{document}

\maketitle

\begin{abstract}
Perfect fluid equations are formulated which are invariant
under the $\ell$-conformal Newton-Hooke group for
an arbitrary integer or half-integer value of the parameter $\ell$.
For $\ell=\frac32$ the corresponding conserved charges are constructed
and the Hamiltonian formulation is built.

\end{abstract}

\thispagestyle{empty}
\newpage
\setcounter{page}{1}

\allowdisplaybreaks

\section{Introduction}\label{Sec1}

Fluid mechanics provides successful macroscopic
description of underlying involved microscopic processes. Such an
effective description applied to strongly coupled systems stimulates
current interest to fluid dynamics with conformal
symmetries. They are in the focus of the fluid/gravity correspondence
\cite{Ran09}. In the case of strongly coupled condensed matter
systems \cite{Son08,BM08,NS2010}, fluid models invariant under
the action of non-relativistic
conformal groups are of interest.

The symmetry group of the Euler equation, which describes the
dynamics of a non-relativistic perfect fluid, might be larger than
the Galilei group. A special choice of the equation of state extends
it to the Schrodinger group \cite{HH1,RS00,JNPP04,HZ}. In addition
to the Galilei transformations, the latter contains dilatation and
special conformal transformations. As is well known, the Galilei
algebra can be considered as a contraction of the Newton-Hooke
algebra \cite{BL67,ABCP99,GP03} in which the cosmological constant
tends to zero (the flat space limit). The Newton-Hooke algebra
follows from the (anti) de Sitter algebra in the non-relativistic
limit in much the same way as the Galilei algebra results from the
Poincar\'e algebra. The principal difference between the Galilei and
Newton-Hooke algebras is that in the latter case the commutator
between the generators of the temporal and spatial translations
yields the Galilei boost $[H,P^{}_i]=\pm\frac{1}{R^2}C_i^{}$, where
$R$ is the characteristic time. In physics literature,
$\Lambda=\pm\frac{1}{c^2 R^2}$, where $c$ is the speed of light, is
identified with the cosmological constant. A natural question arises
as to how to formulate perfect fluid equations in non-relativistic
spacetime with cosmological constant.

One possible way to tackle the problem is to analyze the
non-relativistic limit of the relativistic
hydrodynamics equations formulated in (anti) de Sitter space
\cite{TGHXZ05}. An alternative possibility, which is one of the subjects
of the present paper, is to start with the
non-relativistic hydrodynamics
equations and accommodate the Newton-Hooke symmetry there.

It has been known for a long time that, like the Galilei algebra,
the Newton-Hooke algebra admits a conformal extension which is
parameterized by an integer or half-integer parameter $\ell$
\cite{NOR97}. Its dynamical realizations have been extensively
studied in the past (see e.g.
\cite{DH09,Gal09,Gal10,DH11,GM11,GM13a,AGGM,M14,And14,GM15a,KLS16}
and references therein). In addition to the Newton-Hooke
transformations, the algebra includes dilatation, special conformal
transformation and $2\ell-1$ vector generators associated with the
so-called constant accelerations. When the cosmological constant
tends to zero, it reduces to the $\ell$-conformal Galilei algebra
\cite{GM11}. Note that the instance of $\ell=\frac12$ is relevant
for the harmonic oscillator \cite{N73}. An example of a dynamical
system that accommodates the conformal Newton-Hooke symmetry for
$\ell>\frac12$ is the Pais-Uhlenbeck oscillator \cite{PU}, provided
its frequencies satisfy a special restriction \cite{AGGM}.

Perfect fluid equations with the $\ell$-conformal Galilei
symmetry have been recently constructed in
\cite{Gal22a} and further studied in \cite{Gal22b,Sne23a,Sne24a,Sne24}.
They include the continuity equation, the generalized Euler equation
with higher derivatives and a specific equation of state. The
principal objective of this work is to extend the analysis in
\cite{Gal22a} to include a cosmological constant.

The work is organized as follows.

In the next section, symmetries of the non-relativistic perfect
fluid equations are discussed. First we review symmetries of a
perfect fluid in the absence of external fields. Then external
harmonic potential is added. As shown below, in the latter case the
$\ell=\frac12$ conformal Newton-Hooke symmetry is realized.

In Section \ref{Sec3}, the structure of the $\ell$-conformal
Newton-Hooke algebra is briefly reminded.

In Section \ref{Sec4}, perfect fluid equations are constructed which
hold invariant under the $\ell=\frac32$ conformal Newton-Hooke
group. The Hamiltonian formulation is built and a complete list of
conserved charges is given.

In Section \ref{Sec5}, it is shown that the same equations can be
obtained by applying Niederer's transformation \cite{N73} to the
equations in \cite{Gal22a}. As a by-product, perfect fluid equations
which accommodate the $\ell$-conformal Newton-Hooke symmetry group
for arbitrary integer and half-integer value of the parameter $\ell$
are found.

We summarize our results and discuss possible further developments
in the concluding Section 6.

\section{Symmetries of perfect fluid equations}\label{Sec2}

\subsection{Free perfect fluid equations}\label{SubSec21}

In a non-relativistic space-time with a
temporal coordinate $t$ and spatial coordinates $x_i$, $i=1,...,d$,
a perfect fluid
is characterized by the density $\rho(t,x)$ and the
velocity vector field $\upsilon_i(t,x)$. The evolution over time
is described by the
continuity equation and the Euler equation\footnote{Throughout the
text we use the notations: $\partial_0=\frac{\partial}{\partial t}$,
$\partial_i=\frac{\partial}{\partial x_i}$, ${\cal
D}=\partial_0+\upsilon_i\partial_i$. Summation over repeated indices
is understood. Considering the coordinates $t$ and $x_i$ as
independent we have the identity ${\cal D}x_i=\upsilon_i$.}
\begin{eqnarray}\label{PFEq}
{\partial_0\rho}+ {\partial_i (\rho\upsilon_i)}=0,\quad {\cal
D}\upsilon_i=-\frac{1}{\rho}{\partial_i p}+\frac{f_i}{\rho},
\end{eqnarray}
where $p(t,x)$ is the pressure, which is assumed to be related to
the density via an equation of state $p=p(\rho)$,
and $f_i$ designate external forces.

For a specific equation of state and $f_i=0$ the symmetry group of
(\ref{PFEq}) coincides with
is the Schrodinger group
\cite{RS00}, which in addition to the Galilei transformations includes
dilation and special conformal transformations. One way to see this
is to make recourse to the
non-relativistic energy-momentum tensor (see e.g. \cite{JNPP04})
\begin{align}
&T^{00}=\frac12\rho\upsilon_i\upsilon_i+V, &&
T^{i0}=\rho\upsilon_i(\frac12\upsilon_j\upsilon_j+V')\nn
\\
&T^{0i}=\rho\upsilon_i, &&
T^{ji}=\rho\upsilon_i\upsilon_j+\delta_{ij}p,
\end{align}
where the potential function $V(\rho)$ is related to the pressure
via the Legendre transformations $p=\rho V'-V$. The components
$T^{00}$ and $T^{0i}$ are identified with the energy density and the
energy flux density whereas $T^{i0}$ and $T^{ji}$ link to the
momentum density and the stress tensor. They satisfy the continuity
equations
\begin{eqnarray}\label{Tdifeq}
{\partial_0 T^{00}}+{\partial_i T^{i0}}=0,\qquad {\partial_0
T^{0i}}+{\partial_j T^{ji}}=0,
\end{eqnarray}
as well as the algebraic condition
\begin{eqnarray}\label{Talgeq}
2T^{00}=\delta_{ij}T^{ij},\qquad V=\frac12dp.
\end{eqnarray}
Two comments are in order. Firstly,
$T^{i0}\neq T^{0i}$ because the theory is not
Lorentz-invariant but $T^{ij}= T^{ji}$ because it is invariant under
spatial rotations. Secondly, the condition (\ref{Talgeq}) is
satisfied only for $p\sim\rho^{1+\frac{2}{d}}$, where $d$ is
the spatial dimension, which is the
analogue of the tracelessness condition characterizing a
relativistic conformal field theory.

The continuity equations (\ref{Tdifeq}), the condition
(\ref{Talgeq}) and the properties of the energy-momentum tensor
allow one to construct integrals of motion that correspond to
symmetries of the theory. Denoting conserved charges associated with
the temporal
translation, spatial translation, spatial rotations, Galilei boost,
dilation and special conformal transformation by $H$, $P_i$,
$M_{ij}$, $C_i$, $D$, and $K$, respectively, one readily finds
\begin{eqnarray}
H&=&\int dx T^{00}=\int
dx(\frac12\rho\upsilon_i\upsilon_i+V),\label{HamPF}
\\
P_i^{}&=&\int dx T^{0i}=\int dx \rho \upsilon^{}_i,
\\
C_i^{}&=&\int dx (T^{0i}t-\rho x_i)=tP_i-\int dx \rho x_i,
\\
M_{ij}&=&\int dx (T^{0i}x_j-T^{0j}x_i)=\int dx
(\rho\upsilon_ix_j-\rho\upsilon_jx_i),
\\
D&=&\int dx (T^{00}t-\frac12 T^{0i}x_i)=tH-\frac12\int
dx\rho\upsilon_i^{}x_i,
\\
K&=&\int dx (T^{00}t^2- T^{0i}tx_i+\frac12\rho x_ix_i)=
-t^2H+2tD+\frac12\int dx\rho x_ix_i.
\end{eqnarray}
In order to verify the conservation of $C_i$ and $K$ over time,
one should also use the
continuity equation for the density
$\partial_0\rho+\partial_iT^{0i}=0$.

Within the Hamiltonian formulation \cite{MG80}, which is
defined by the Hamiltonian
(\ref{HamPF}) and the Poisson brackets
\begin{eqnarray}\label{PBPF}
\{\rho(x),\upsilon_i(y)\}=- {\partial_i}\delta(x-y),\quad
\{\upsilon_i(x),\upsilon_j(y)\}=\frac{1}{\rho}\left({\partial_i\upsilon_j}
-{\partial_j\upsilon_i}\right)\delta(x-y),
\end{eqnarray}
the conserved charges do satisfy the structure relations of
the Schrodinger
algebra.

\subsection{Perfect fluid equations in the harmonic trap}\label{SubSec22}

Let us consider a perfect fluid in the harmonic trap specified by
$f_i=-\omega^2\rho x_i$, where $\omega^2$ is
a positive constant of dimension $[\omega]=t^{-1}$, which
is assumed to be small.
The Euler equation in (\ref{PFEq}) takes on the form
\begin{eqnarray}\label{PFEqNH} {\cal
D}\upsilon_i+\omega^2x_i=-\frac{1}{\rho}{\partial_i p}.
\end{eqnarray}

Together with the continuity equation the equation (\ref{PFEqNH})
can be represented in the Hamiltonian form
\begin{eqnarray}
\partial_0\rho=\{\rho,H\}=-\partial_i(\rho\upsilon_i),\quad
\partial_0\upsilon_i=\{\upsilon_i,H\}=-\upsilon_j\partial_j\upsilon_i-\omega^2x_i-\frac{1}{\rho}{\partial_i p}
\end{eqnarray}
where
\begin{eqnarray}
H=\frac12\rho\upsilon^{}_i\upsilon^{}_i+\frac12\omega^2 \rho
x_ix_i+V,\quad p=\rho V'-V,
\end{eqnarray}
and the Poisson brackets are specified in (\ref{PBPF}).

Similarly to the harmonic oscillator \cite{GP03},
one can construct
integrals of motion that link to spatial translations, the Galilei boost
and spatial rotations
\begin{eqnarray}\label{Hamdl}
P_i^{}&=&\int dx (\rho
\upsilon^{}_i\cos{\omega t}+\omega\rho x_i\sin{\omega t} ),\\
C_i^{}&=&\frac{1}{\omega}\int dx (\rho \upsilon^{}_i\sin{\omega
t}-\omega\rho x_i\cos{\omega t} ),\\
M_{ij}&=&\int dx (\rho\upsilon_ix_j-\rho\upsilon_jx_i),
\end{eqnarray}
which jointly with $H$ and satisfy the following structure
relations with respect to the Poisson bracket
\begin{align}\label{NHalg}
&  \{H,P^{}_i\}=-\frac{1}{R^2}C_i^{}, &&
\{P_i,M_{jk}^{}\}=\delta_{ij}P_{k}-\delta_{ik}P_{j},\nn
\\
& \{H,C^{}_i\}=P_i, &&
\{C_i,M_{jk}^{}\}=\delta_{ij}C_{k}-\delta_{ik}C_{j},\nn
\\
& \{P_i^{},C_j^{}\}=\delta_{ij} m, &&
\{M_{ij}^{},M_{ab}^{}\}=\delta_{i[a}M_{b]j}-\delta_{j[a}M_{b]i},
\end{align}
where we identified $\omega^2=\frac{1}{R^2}$. The relations
(\ref{NHalg}) define the Newton-Hooke algebra \cite{BL67} with a
negative cosmological constant\footnote{The case of a positive
cosmological constant is obtained by a formal replacement
$\omega\rightarrow i\omega$.} $\Lambda=-\frac{1}{R^2}$, extended by
the central charge $m=\int dx\rho$.

Like the Galilei algebra, the Newton-Hooke algebra admits a
conformal extension \cite{NOR97} by the generators of dilatation $D$
and special conformal transformation $K$. Additional structure
relations read \cite{Gal10}
\begin{align}\label{confNH}
& {[H,D]}=H\mp\frac{2}{R^2}K, && {[D,P^{}_i]}=-\frac12P^{}_i,\nn
\\
& {[H,K]}=2D, && {[D,C^{}_i]}=\frac12C^{}_i,\nn
\\
& {[D,K]}=K, && {[K,P^{}_i]}=-C^{}_i,
\end{align}
where the upper/lower sign in the commutator $[H,D]$ corresponds to
the negative/positive cosmological constant.

Let us construct conserved charges that realize extra conformal symmetries for
the perfect fluid model under consideration. As in the free case, it
seems natural to search for them in the quadratic
form
\begin{eqnarray}\label{Hamdl}
J&=&\int
dx\Large(\beta_1(t)\rho\upsilon^{}_i\upsilon^{}_i+\beta_2(t)\rho\upsilon_ix_i+\beta_3(t)\rho
x_ix_i+\beta_4(t)V\Large),
\end{eqnarray}
where we added a term with potential $V$ and introduced arbitrary
coefficients $\beta_i$ that depend only on time. From the condition
$\partial_0 J=0$ a system of equations arises (the dot denotes the time
derivative)
\begin{eqnarray}\label{Hamdl}
\dot\beta_1+\beta_2=0,\quad
\dot\beta_2+2(\beta_3-\beta_1\omega^2)=0,\quad
\dot\beta_3-\beta_2\omega^2=0,\quad 2\beta_1-\beta_4=0,
\end{eqnarray}
and the same condition on the potential $V=\frac12dp$, where $d$ is the spatial dimension,
as in the free
case (\ref{Talgeq}). The general solution is easily found
\begin{eqnarray}\label{Hamdl}
\beta_1&=&\frac12\beta_4=c_1+c_2\cos{2\omega t}+c_3\sin{2\omega
t},\nn
\\
\beta_2&=&2\omega(c_2\sin{2\omega t}-c_3\cos{2\omega t}),\nn
\\
\beta_3&=&\omega^2(c_1-c_2\cos{2\omega t}-c_3\sin{2\omega t}),
\end{eqnarray}
which
contains three arbitrary constants $c_{1,2,3}$ meaning that there are three
independent integrals of motion. As independent integrals we choose
$J_i=J|_{c_i=\frac12,c_{j\neq i}=0}$, which yield
\begin{eqnarray}
J_1&=&\int
dx(\frac12\rho\upsilon^{}_i\upsilon^{}_i+\frac12\omega^2\rho
x_ix_i+V)=H,\nn
\\
J_2&=&H\cos{2\omega t}+\omega\int dx(\rho\upsilon_ix_i\sin{2\omega
t}-\omega\rho x_ix_i\cos{2\omega t}),\nn
\\
J_3&=&H\sin{2\omega t}-\omega\int dx(\rho\upsilon_ix_i\cos{2\omega
t}+\omega\rho x_ix_i\sin{2\omega t}).
\end{eqnarray}
The first integral of motion corresponds to the previously obtained
expression for the total energy $J_1=H$ while other two should be related
to $D$ and $K$. Computing
the Poisson brackets
\begin{eqnarray}
&\{J_1,J_2\}=-2\omega J_3,\quad \{J_1,J_3\}=2\omega J_2,\quad
\{J_2,J_3\}=2\omega J_1,\nn
\\
& \{P_i^{},J_2\}=-\omega^2 C_i^{},\quad \{C_i^{},J_2\}=-
P_i^{},\quad \{P_i^{},J_3\}=\omega P_i^{},\quad
\{C_i^{},J_3\}=-\omega C_i^{},
\end{eqnarray}
and taking into account (\ref{confNH}), one finally gets
\begin{eqnarray}
D=\frac{1}{2\omega} J_3,\quad K=\frac{1}{2\omega^2}(J_1- J_2),\quad
\omega^2=\frac{1}{R^2}.
\end{eqnarray}

To summarize, the generalized Euler equations (\ref{PFEqNH})
enjoy the conformal Newton-Hooke
symmetry (with negative cosmological constant) (\ref{confNH})
provided the equation of
state $p\sim\rho^{1+\frac{2}{d}}$ is chosen as in the flat case.

\section{The $\ell$-conformal Newton-Hooke  algebra }\label{Sec3}

In the previous section, we established the conformal Newton-Hooke
symmetry of the perfect fluid equations in the harmonic trap.
Such a conformal extension of the Newton-Hooke algebra is
not unique. There is a one-parameter family of finite-dimensional
conformal extensions \cite{NOR97,GM11}
\begin{align}\label{lconfNH}
& {[H,D]}=H\mp\frac{2}{R^2}K, &&
{[H,C^{(k)}_i]}=kC^{(k-1)}_i\pm\frac{(k-2\ell)}{R^2}C_i^{(k+1)},\nn
\\
& {[H,K]}=2D, && {[D,C^{(k)}_i]}=(k-\ell)C^{(k)}_i,\nn
\\
& {[D,K]}=K, && {[K,C^{(k)}_i]}={(k-2\ell)}C^{(k+1)}_i,\nn
\\
& {[C_i^{(k)},M_{ab}]}=\delta_{ia}C_{b}^{(k)}-\delta_{ib}C_a^{(k)},
&& {[M_{ij},M_{ab}]}=\delta_{i[a}M_{b]j}-\delta_{j[a}M_{b]i},
\end{align}
where $k=0,1,...,2\ell$ and the parameter $\ell$ is an arbitrary
integer or half-integer number. Generators $H$, $D$, $K$, $M_{ij}$
correspond to time translation, dilation, special conformal
transformation, spatial rotations, while the vector generators $C^{(k)}_i$
correspond to spatial translation and Galilei boost for $k=0,1$ and constant
accelerations for $k>1$. As above, a real constant $R$ is the
characteristic time which links to the negative/positive
cosmological constant $\Lambda=\mp\frac{1}{c^2 R^2}$, where $c$ is the speed of light.

In the non-relativistic space-time $(t,x_i)$ the algebra
(\ref{lconfNH}) with negative cosmological constant can be realized
as follows \cite{GM11}
\begin{eqnarray}\label{RealNH}
&&H={\partial}_0,\quad
D=\frac12R\left(\sin{\frac{2t}{R}}\right){\partial}_0+\ell\left(\cos{\frac{2t}{R}}\right)
x_i{\partial}_i,\nn\\
&& K=\frac12R^2\left(1-\cos{\frac{2t}{R}}\right){\partial}_0+\ell
R\left(\sin{\frac{2t}{R}}\right)x_i{\partial}_i,\nn\\
&&
C^{(k)}_i=R^k\left(\tan{\frac{t}{R}}\right)^k\left(\cos{\frac{t}{R}}\right)^{2\ell}{\partial}_i,\quad
M_{ij}=x_i\partial_j-x_j\partial_i,
\end{eqnarray}
while the case with a positive cosmological constant is obtained by
the formal replacement $R\rightarrow iR$.

In arbitrary dimension and for half-integer $\ell$, conformal
Newton-Hooke algebra admits a central extension \cite{GM11}
\begin{eqnarray}\label{CentNH}
[C_i^{(k)},C_j^{(m)}]&=&(-1)^{k}k!m!\delta_{(k+m)(2\ell)}\delta_{ij}m,
\end{eqnarray}
where the central charge $m$ links to mass in dynamical realizations.

Note that making a linear change of the basis $H\rightarrow
H\mp\frac{1}{R}K$ in (\ref{lconfNH}), one reproduces the
$\ell$-conformal Galilei algbera. However, they are usually treated
separately because a change of the Hamiltonian alters the dynamics.
Notice also that the Newton-Hooke case is characterized by a
dimensionfull constant $R$, which is absent in the case of the
$\ell$-conformal Galilei algbera.

\section{Perfect fluid with the $\ell$-conformal Newton-Hooke symmetry}\label{Sec4}

Bearing in mind that the $\ell$-conformal Newton-Hooke algebra is the
cosmological extension of the $\ell$-conformal Galilei algebra,
we begin
with the perfect fluid equations realizing the latter symmetry group
\cite{Gal22a}
\begin{eqnarray}
&&{\partial_0\rho}+ {\partial_i (\rho\upsilon_i)}=0,\label{PFEqlCG1}
\\
&&{\cal D}^{2\ell}\upsilon_i=-\frac{1}{\rho}{\partial_i
p},\label{PFEqlCG2}
\\
&& p=\nu\rho^{1+\frac{1}{\ell d}},\label{PFEqlCG3}
\end{eqnarray}
where $\nu$ is a constant. Their invariance under transformations
from the $\ell$-conformal Galilei group was explicitly shown in
\cite{Gal22a} for an arbitrary integer or half-integer $\ell$.
Alternatively, for a half-integer $\ell$ one can go over to the
Hamiltonian formulation and establish the algebra with the use of
the Poisson bracket \cite{Sne23a}. The equations above contain the
continuity equation for the density (\ref{PFEqlCG1}), the
generalized Euler equation with higher derivatives (\ref{PFEqlCG2})
and the equation of state (\ref{PFEqlCG3}). For $\ell=\frac12$ they
correctly reproduce the perfect fluid equations with Schrodinger
symmetry \cite{JNPP04}.

In order to accommodate the $\ell$-conformal Newton-Hooke group,
it appears natural to
deform only the generalized Euler equation and leave the continuity
equation and the equation of state unchanged. Focusing in what follows on
the case of $\ell=\frac32$ we modify the generalized third-order Euler
equation as follows
\begin{eqnarray}\label{EulerEqCNH32}
 {\cal
D}^{3}\upsilon_i+(\omega_1^2+\omega_2^2){\cal
D}\upsilon_i+\omega_1^2\omega_2^2x_i=-\frac{1}{\rho}{\partial_i p},
\end{eqnarray}
where we added a term with a single derivative and a harmonic potential
term introducing two arbitrary parameters $\omega_2^2>\omega_1^2>0$
of dimension $[\omega_1]=[\omega_2]=t^{-1}$. With this choice of
the parameters, the left-hand side of the equation (\ref{EulerEqCNH32})
is an analogue of the Pais-Uhlenbeck oscillator \cite{PU} in
classical mechanics.

Introducing the Ostrogratsky-like auxiliary field variables
$\upsilon_i^{0},\upsilon_i^{1},\upsilon_i^{2}$ with
$\upsilon_i^{0}=\upsilon_i$ the equation (\ref{EulerEqCNH32}) can be
derived from the Hamiltonian
\begin{eqnarray}\label{HamlCNH}
H&=&\int dx\left[\rho\left(\upsilon^{0}_i\upsilon^{2}_i
-\frac12\upsilon^{1}_i\upsilon^{1}_i-\frac12(\omega^2_1+\omega^2_2)\upsilon^{0}_i\upsilon^{0}_i+
\frac12\omega^2_1\omega^2_2 x_ix_i\right)+V\right],
\end{eqnarray}
where the potential $V$ links to the pressure via the Legendre transform
$p=\rho V'-V$, provided the Poisson brackets
\begin{eqnarray}\label{PBFluidlCNH}
&\{\rho(x),\upsilon^{2}_i(y\}=-{\partial_i}\delta(x-y),\quad
\{\upsilon^{0}_i(x),\upsilon^{2}_j(y)\}=-\frac{1}{\rho} {\partial_j
\upsilon^{0}_i}\delta(x-y),\nn
\\
&\{\upsilon^{0}_i(x),\upsilon^{1}_j(y)\}=-\frac{1}{\rho}\delta_{ij}\delta(x-y),\quad
\{\upsilon^{1}_i(x),\upsilon^{2}_j(y)\}=-\frac{1}{\rho} {\partial_j
\upsilon^{1}_i}\delta(x-y),\nn
\\
&\{\upsilon^{2}_i(x),\upsilon^{2}_j(y)\}=\frac{1}{\rho}\left({\partial_i
\upsilon^{2}_j}- {\partial_j \upsilon^{2}_i}\right)\delta(x-y),
\end{eqnarray}
are used. Indeed, the dynamical
equations have the form
\begin{eqnarray}\label{HamPFeq}
&&\partial_0\rho=\{\rho,H\}=-\partial_i(\rho\upsilon_i^0),\nn
\\
&&\partial_0\upsilon^0_i=\{\upsilon^0_i,H\}=-\upsilon^0_j\partial_j\upsilon^0_i+\upsilon_i^{1},\nn
\\
&&\partial_0\upsilon^1_i=\{\upsilon^1_i,H\}=-\upsilon^0_j\partial_j\upsilon^1_i-(\omega^2_1+\omega^2_2)\upsilon_i^{0}+\upsilon_i^{2},\nn
\\
&&\partial_0\upsilon^2_i=\{\upsilon^2_i,H\}=-\upsilon^0_j\partial_j\upsilon^2_i-\omega^2_1\omega^2_2
x_i-\partial_iV',
\end{eqnarray}
the first of which gives the continuity equation for density. Eliminating
the auxiliary variables $\upsilon_i^{1},\upsilon_i^{2}$
from the second and third equations and substituting them into the
fourth equation, the generalized Euler equation (\ref{EulerEqCNH32}) is
reproduced.

Note that the non-canonical Poisson brackets
(\ref{PBFluidlCNH}) are the same as those for the undeformed theory
(\ref{PFEqlCG2}) originally introduced in\cite{Sne23a}.

As the next step, let us construct the corresponding conserved charges.
We start with vector generators
$C_i^{(0)},C_i^{(1)},C_i^{(2)},C_i^{(3)}$ and choose them as
linear expressions in the field variables
$\upsilon^{0}_i,\upsilon^{1}_i,\upsilon^{2}_i$ and spatial
coordinate $x_i$ multiplied by the density $\rho$. In general, a
conserved charge can depend explicitly on time so the most general
expression reads
\begin{eqnarray}
I_i=\int dx\big(\alpha_1(t) \rho \upsilon^{2}_i+\alpha_2(t)\rho
\upsilon^{1}_i+\alpha_3(t) \rho \upsilon^{0}_i+\alpha_4(t) \rho
x_i\big),
\end{eqnarray}
where $\alpha_i$ are arbitrary time-depended coefficients. The
conservation condition $\partial_0 I_i=0$ gives a system of
differential equations
\begin{eqnarray}\label{Hamdl}
\dot\alpha_1+\alpha_2=0,\quad \dot\alpha_2+\alpha_3=0,\quad
\dot\alpha_3+\alpha_4-(\omega^2_1+\omega^2_2)\alpha_2=0,\quad
\dot\alpha_4-\alpha_1\omega^2_1\omega^2_2=0
\end{eqnarray}
which has the general solution
\begin{eqnarray}\label{Hamdl}
\alpha_1&=&c_1\cos{\omega_1 t}+c_2\sin{\omega_1 t}+c_3\cos{\omega_2
t}+c_4\sin{\omega_2 t},\nn
\\
\alpha_2&=&c_1\omega_1\sin{\omega_1 t}-c_2\omega_1\cos{\omega_1
t}+c_3\omega_2\sin{\omega_2 t}-c_4\omega_2\cos{\omega_2 t},\nn
\\
\alpha_3&=&-c_1\omega_1^2\cos{\omega_1 t}-c_2\omega_1^2\sin{\omega_1
t}-c_3\omega_2^2\cos{\omega_2 t}-c_4\omega_2^2\sin{\omega_2 t},\nn
\\
\alpha_4&=&c_1\omega_1\omega^2_2\sin{\omega_1
t}-c_2\omega_1\omega^2_2\cos{\omega_1
t}+c_3\omega_2\omega^2_1\sin{\omega_2
t}-c_4\omega_2\omega^2_1\cos{\omega_2 t}.
\end{eqnarray}
It is satisfied for arbitrary $\omega_2^2>\omega_1^2$ and contains
four integration constants $c_{1,2,3,4}$ such that there are four
functionally independent integrals of motion. For simplicity we
choose them in the form in which three constants are zero and the
fourth constant is equal to one
\begin{eqnarray}
I_i^{1}&=&\int dx(\cos{\omega_1 t}\rho
\upsilon^{2}_i+\omega_1\sin{\omega_1 t}\rho
\upsilon^{1}_i-\omega_1^2\cos{\omega_1 t}\rho
\upsilon^{0}_i+\omega_1\omega^2_2\sin{\omega_1 t}\rho
x_i),\nn\\
I_i^2&=&\int dx(\sin{\omega_1 t}\rho
\upsilon^{2}_i-\omega_1\cos{\omega_1 t}\rho
\upsilon^{1}_i-\omega_1^2\sin{\omega_1 t}\rho
\upsilon^{0}_i-\omega_1\omega^2_2\cos{\omega_1 t}\rho x_i),\nn
\\
I_i^3&=&\int dx(\cos{\omega_2 t}\rho
\upsilon^{2}_i+\omega_2\sin{\omega_2 t}\rho
\upsilon^{1}_i-\omega_2^2\cos{\omega_2 t}\rho
\upsilon^{0}_i+\omega_2\omega^2_1\sin{\omega_2 t}\rho
x_i),\nn\\
I_i^4&=&\int dx(\sin{\omega_2 t}\rho
\upsilon^{2}_i-\omega_2\cos{\omega_2 t}\rho
\upsilon^{1}_i-\omega_2^2\sin{\omega_2 t}\rho
\upsilon^{0}_i-\omega_2\omega^2_1\cos{\omega_2 t}\rho x_i).
\end{eqnarray}
We will establish the explicit relation of these four integrals
of motion to the vector generators $C_i^{(k)}$ at the end of the
section. Here we only write down the brackets among
$(I_i^1,I_i^2,I_i^3,I_i^4)$ and $H$
\begin{align}\label{ComCond1}
& \{I_i^1,H\}=\omega_1I_i^2, && \{I_i^3,H\}=\omega_2I_i^4, &&
\{I_i^1,I_j^2\}=\omega_1(\omega_2^2-\omega_1^2)m\delta_{ij},\nn
\\
& \{I_i^2,H\}=-\omega_1I_i^1, && \{I_i^4,H\}=-\omega_2I_i^3, &&
\{I_i^3,I_j^4\}=-\omega_2(\omega_2^2-\omega_1^2)m\delta_{ij},
\end{align}
where $m=\int dx \rho$ is the conserved total mass.

Let us turn to the construction of conserved charges associated with
the dilatation $D$ and special conformal transformation $K$. We
search for them as quadratic combinations involving
$\upsilon^{0}_i,\upsilon^{1}_i,\upsilon^{2}_i$ and $x_i$ multiplied
by the density $\rho$. The most general expression with arbitrary
time-dependent coefficients $\beta_i$ reads
\begin{eqnarray}
J&=&\int dx\left(\beta_1(t)\rho\upsilon^{0}_i\upsilon^{2}_i
+\beta_2(t)\rho\upsilon^{1}_i\upsilon^{1}_i+\beta_3(t)\rho\upsilon^{2}_ix_i+\beta_4(t)\rho\upsilon^{1}_i\upsilon^{0}_i\right.\nn\\
&&\left.+\beta_5(t)\rho\upsilon^{0}_i\upsilon^{0}_i+\beta_6(t)\rho\upsilon^{1}_ix_i+\beta_7(t)\rho\upsilon^{0}_ix_i
+\beta_8(t) \rho x_ix_i+\beta_9(t)V\right),
\end{eqnarray}
where we also included a term with the potential $V$. From the
conservation condition $\partial_0J=0$ one obtains the restrictions
\begin{align}
& \beta_1+2\beta_2=0, && \beta_1-\beta_9=0,  &&
\dot\beta_4-2\beta_2(\omega^2_1+\omega^2_2)+2\beta_5+\beta_6=0,\nn
\\
& \dot\beta_1+\beta_3+\beta_4=0, && \dot\beta_6+\beta_7=0,  &&
\dot\beta_5-\beta_4(\omega^2_1+\omega^2_2)+\beta_7=0,\nn
\\
& \dot\beta_2+\beta_4=0, &&
\dot\beta_8-\beta_3\omega^2_1\omega^2_2=0, &&
\dot\beta_7-\beta_1\omega^2_1\omega^2_2-\beta_6(\omega^2_1+\omega^2_2)+2\beta_8=0,\nn
\\
& \dot\beta_3+\beta_6=0, && \beta'_9V+\beta_3dp=0,&&
\end{align}
which prove compatible provided the extra restrictions
\begin{eqnarray}\label{NHcond}
\omega_2=3\omega_1,\quad V=\frac32dp
\end{eqnarray}
are imposed. Then the coefficients $\beta$ acquire the form
\begin{eqnarray}
\beta_1&=&-2\beta_2=\beta_9=c_1+c_2\cos{2\omega_1 t}
+c_3\sin{2\omega_1 t},\nn
\\
\beta_4&=&-\frac13\beta_3=-\omega_1(c_2 \sin{2\omega_1 t}
-c_3\cos{2\omega_1 t}),\nn
\\
\beta_5&=&-\omega^2_1(5c_1+c_2\cos{2\omega_1 t} +c_3\sin{2\omega_1
t}),\nn
\\
\beta_6&=&-6\omega_1^2(c_2\cos{2\omega_1 t} +c_3\sin{2\omega_1
t}),\nn
\\
\beta_7&=&-12\omega_1^3 (c_2\sin{2\omega_1 t} -c_3\cos{2\omega_1
t}),\nn
\\
\beta_8&=&\frac{9\omega^4_1}{2}({c_1} -3{c_2}\cos{\omega t}
-3c_3\sin{\omega t}),
\end{eqnarray}
which contain three constants of integration $c_{1,2,3}$. As
three independent integrals of motion we choose those obtained by
setting two constants to vanish and equating the last one to unity
\begin{eqnarray}
J_1&=&\int dx\left[\rho\left(\upsilon^{0}_i\upsilon^{2}_i
-\frac12\upsilon^{1}_i\upsilon^{1}_i-5\omega^2_1\upsilon^{0}_i\upsilon^{0}_i+
\frac92\omega^4_1 x_ix_i\right)+V\right]=H,\nn
\\
J_2&=&\cos{2\omega_1 t}H+\int dx\rho\left[\omega_1 \sin{2\omega_1
t}(3\upsilon^{2}_ix_i-\upsilon^{1}_i\upsilon^{0}_i-12\omega_1^2
\upsilon^{0}_ix_i)\right.\nn\\
&&\left.+2\omega_1^2\cos{2\omega_1
t}(2\upsilon^{0}_i\upsilon^{0}_i-3\upsilon^{1}_ix_i -{9\omega^2_1}
x_ix_i)\right],\nn
\\
J_3&=&\sin{2\omega_1 t}H-\int dx\rho\left[\omega_1 \cos{2\omega_1
t}(3\upsilon^{2}_ix_i-\upsilon^{1}_i\upsilon^{0}_i
-12\omega_1^2 \upsilon^{0}_ix_i)\right.\nn\\
&&\left.-2\omega_1^2\sin{2\omega_1
t}(2\upsilon^{0}_i\upsilon^{0}_i-3\upsilon^{1}_ix_i -{9\omega^2_1}
x_ix_i)\right].
\end{eqnarray}

Then it is straightforward to establish the following structure relations
\begin{eqnarray}\label{ComCond2}
\{J_1,J_2\}&=&-2\omega_1 J_3,\quad \{J_3,J_1\}=-2\omega_1 J_2,\quad
\{J_2,J_3\}=2\omega_1 J_1
\end{eqnarray}
and
\begin{align}\label{ComCond3}
& \{J_2,I_i^1\}=2\omega_1I_i^2+\omega_1I_i^4, &&
\{J_2,I_i^3\}=-3\omega_1I_i^2,\nn
\\
& \{J_2,I_i^2\}=2\omega_1I_i^1-\omega_1I_i^3, &&
\{J_2,I_i^4\}=3\omega_1I_i^1,\nn
\\
& \{J_3,I_i^1\}=-2\omega_1I_i^1-\omega_1I_i^3, &&
\{J_3,I_i^3\}=-3\omega_1I_i^1,\nn
\\
& \{J_3,I_i^2\}=2\omega_1I_i^2-\omega_1I_i^4, &&
\{J_3,I_i^4\}=-3\omega_1I_i^2.
\end{align}
Comparing the relations above, as well as (\ref{ComCond1}), to the
structure relations of the $\ell=\frac32$ conformal Newton-Hooke
algebra (\ref{lconfNH}) and (\ref{CentNH}), one finds the desired identifications
\begin{align}\label{ConsChar}
& D=\frac{1}{2\omega_1} J_3, && C_i^{(0)}=\frac14(3I_i^1+I_i^3), &&
C_i^{(2)}=\frac{1}{4\omega_1^2}(I_i^1-I_i^3),\nn
\\
& K=\frac{1}{2\omega_1^2}(H-J_2), &&
C_i^{(1)}=\frac{1}{4\omega_1}(I_i^2+I_i^4), &&
C_i^{(3)}=\frac{1}{4\omega_1^3}(3I_i^2-I_i^4),
\end{align}
with $\omega_1^2=\frac{1}{R^2}$ for the case of
negative cosmological constant. In the limit $\omega_1\rightarrow 0$
the conserved charges
(\ref{ConsChar})
reproduce those corresponding to the
$\ell=\frac32$ conformal Galilei algebra in \cite{Sne23a}.

To complete analysis, we must also add the conserved charges associated with
spatial rotations
\begin{eqnarray}\label{RotConsCh}
M_{ij}=\int
dx\rho(\upsilon^2_ix_j-\upsilon^2_jx_i+\upsilon^0_i\upsilon^1_j-
\upsilon^0_j\upsilon^1_i).
\end{eqnarray}

Thus we have demonstrated that the generalized perfect fluid equations
(\ref{PFEqlCG1}), (\ref{PFEqlCG3}), (\ref{EulerEqCNH32}) possess the
$\ell=\frac32$-conformal Newton-Hooke
symmetry provided the conditions (\ref{NHcond})
hold. The corresponding conserved charges are determined by
(\ref{ConsChar}) and (\ref{RotConsCh}) and under the
Poisson bracket (\ref{PBFluidlCNH}) they satisfy the algebra
(\ref{lconfNH}). The first condition in (\ref{NHcond}) includes
the constraint on the free parameters $\omega_2=3\omega_1$ which
coincides with the condition on the frequencies for the conformally
invariant Pais-Uhlenbeck oscillator in classical mechanics
\cite{AGGM}. The second condition in (\ref{NHcond}) restricts the form
of the potential $V=\frac32dp$ which is compatible with the equation of
state $p\sim\rho^{1+\frac{2}{3d}}$ as in the flat space
(\ref{PFEqlCG3}).

\section{Niederer's transformation}\label{Sec5}

As was mentioned in Section \ref{Sec3}, the $\ell$-conformal
Newton-Hooke algebra is the conterpart of the $\ell$-conformal
Galilei algebra in the presence of the cosmological constant. The
corresponding realization of the latter reads
\begin{eqnarray}\label{RealCG}
H={\partial}_0,\quad D=t{\partial}_0+\ell x_i{\partial}_i,\quad
K=t^2{\partial}_0+2\ell tx_i{\partial}_i,\quad
C^{(k)}_i=t^k{\partial}_i,
\end{eqnarray}
and can be obtained from (\ref{RealNH}) in the limit $R\rightarrow\infty$.

On the other hand, there exists a coordinate transformation
\cite{GM11} which links \footnote{It is
necessary to take into account the replacement of the basis
$H\rightarrow H\pm\frac{1}{R}K$ in the $\ell$-conformal Galilei
algebra.} (\ref{RealNH}) to (\ref{RealCG})
\begin{eqnarray}\label{NedTrans}
t'=R\tan{\frac{t}{R}},\quad x'_i=\left(\frac{\partial t'}{\partial
t}\right)^\ell x_i=(\cos{\frac{t}{R}})^{-2\ell} x_i,
\end{eqnarray}
where coordinates with prime parameterize the flat space.
For $\ell=\frac12$ these transformations were first introduced by
Niederer in \cite{N73}, where they (locally) link
a free particle to the harmonic
oscillator.

In the previous sections, we constructed perfect fluid equations
with $\ell=\frac{1}{2},\frac{3}{2}$-conformal Newton-Hooke symmetry.
Let us demonstrate that the same results can be obtained from
(\ref{PFEqlCG1}-\ref{PFEqlCG3}) by applying an analogue of the
Niederer transformation (\ref{NedTrans}).

First of all, let us establish how the density and the velocity vector field
are
transformed under (\ref{NedTrans}). The density
transformation is obtained by requiring
the mass of a $d$-dimensional volume element to be invariant
\begin{eqnarray*}
\int_{V'}dx'\rho'(t',x')=\int_{V}dx\rho(t,x),
\end{eqnarray*}
where the measure $dx'=dx'_1...dx'_d$ is transformed as follows
$dx'=|\frac{\partial x'_i}{\partial x_j}|dx$. The result reads
\begin{eqnarray}\label{DenTrans}
\rho'(t',x')&=&({\cos{\frac{t}{R}}})^{2\ell d}\rho(t,x).
\end{eqnarray}

To obtain the transformation law for
$\upsilon_i(t,x)$, consider the orbit of a fluid particle $x_i(t)$
and take into account that
\begin{eqnarray*}
\frac{dx_i(t)}{dt}=\upsilon_i(t,x(t)).
\end{eqnarray*}
Differentiating the second relation in (\ref{NedTrans}), one obtains
\begin{eqnarray}\label{VelTrans}
\upsilon'_i(t',x')&=&({\cos{\frac{t}{R}}})^{-2\ell+2}\left(\upsilon_i(t,x)+\frac{2\ell}{R}{\tan{\frac{t}{R}}}x_i\right).
\end{eqnarray}

Taking into account the identities
\begin{eqnarray*}
\frac{\partial}{\partial t}&=&(\frac{\partial t'}{\partial
t})\frac{\partial}{\partial t'}+(\frac{\partial x'_i}{\partial
t})\frac{\partial}{\partial x'_i},\quad \frac{\partial}{\partial
x_i}=(\frac{\partial t'}{\partial x_i})\frac{\partial}{\partial
t'}+(\frac{\partial x'_j}{\partial x_i})\frac{\partial}{\partial
x'_j},
\end{eqnarray*}
and equations (\ref{DenTrans}), (\ref{VelTrans}), one finds how the
left-hand side of continuity equation is transformed
\begin{eqnarray}\label{PFEqdl}
{\partial'_0\rho'}+ {\partial'_i
(\rho'\upsilon'_i)}&=&({\cos{\frac{t}{R}}})^{2(\ell
d+1)}\left({\partial_0}\rho + {\partial_i}(\rho\upsilon_i)\right),
\end{eqnarray}
so that the continuity equation kept intact.

In order to analyze the generalized
Euler equation (\ref{PFEqlCG2}), one has to establish
how
${\cal D}\upsilon_i$, ${\cal D}^2\upsilon_i$ etc. are
transformed. Taking into account (\ref{VelTrans}) and
\begin{eqnarray}
{\cal D}'=(\cos{\frac{t}{R}})^{2}{\cal D},
\end{eqnarray}
one gets
\begin{eqnarray}
{\cal D}'\upsilon'_i&=&({\cos{\frac{t}{R}}})^{3}({\cal
D}\upsilon_i+\frac{1}{R^2}x_i),\quad {\cal
D}'^2\upsilon'_i=({\cos{\frac{t}{R}}})^{4}({\cal
D}^2+\frac{4}{R^2})\upsilon_i,
\end{eqnarray}
for $\ell=\frac12$ and $\ell=1$. Similarly,
for an arbitrary (half)-integer $\ell$ one
can establishes the relations
\begin{eqnarray*}
{\cal
D}'^{2n-1}\upsilon'_i&=&({\cos{\frac{t}{R}}})^{2n+1}\prod_{k=1}^{n-1}({\cal
D}^2+\frac{(2k+1)^2}{R^2})(D\upsilon_i+\frac{1}{R^2}x_i),\quad
\ell=n-\frac12,
\\
{\cal
D}'^{2n}\upsilon'_i&=&({\cos{\frac{t}{R}}})^{2n+2}\prod_{k=1}^n({\cal
D}^2+\frac{(2k)^2}{R^2})\upsilon_i,\quad \ell=n,
\end{eqnarray*}
where $n=1,2,...$.

The right-hand side of
(\ref{PFEqlCG2}) is transformed as follows
\begin{eqnarray}
-\frac{1}{\rho'}\partial'_ip'&=&-({\cos{\frac{t}{R}}})^{2(\ell+1)}\frac{1}{\rho}\partial_ip,
\end{eqnarray}
where the equation of state $p=\nu\rho^{1+\frac{1}{\ell d}}$
was used.

As a result, after applying (\ref{NedTrans}) to
the generalized Euler equation,
one obtains
\begin{eqnarray}\label{EulerNHhi}
&&\prod_{k=1}^{n-1}({\cal
D}^2+\frac{(2k+1)^2}{R^2})(D\upsilon_i+\frac{1}{R^2}x_i)=-\frac{1}{\rho}\partial_ip
\end{eqnarray}
for a half-integer $\ell=n-\frac12$ and
\begin{eqnarray}\label{EulerNHi}
&&\prod_{k=1}^n({\cal
D}^2+\frac{(2k)^2}{R^2})\upsilon_i=-\frac{1}{\rho}\partial_ip
\end{eqnarray}
for an integer $\ell=n$.

To summarize, the generalized Niederer transformation does not alter the
continuity equation (\ref{PFEqlCG1}) and the equation of
state (\ref{PFEqlCG3}), while it modifies the Euler equation
(\ref{EulerNHhi}) or (\ref{EulerNHi}). By construction,
the equations hold invariant under the $\ell$-conformal Newton-Hooke
transformations and in the particular cases $\ell=\frac12$ and
$\ell=\frac32$ reproduce the results obtained in the previous
sections.

\section{Conclusion}\label{Sec6}

To summarize, in this work we formulated perfect fluid equations
which enjoy the $\ell$-conformal Newton-Hooke symmetry. For
$\ell=\frac12$, the symmetries are naturally realized by the
harmonic trap potential and imposing a suitable equation of state.
For higher values of $\ell$, the symmetries demand a higher
derivative generalization of the Euler equation which is an analogue
of the Pais-Uhlenbeck oscillator in classical mechanics. It was
demonstrated that the same results can be achieved by applying a
generalized Neiderer transformation. For $\ell=\frac32$, the
Hamiltonian formulation was built and the corresponding conserved
charges were constructed.

Turning to possible further developments, it would be interesting to
construct a consistent Lagrangian formulation for perfect fluid
equations with the $\ell$-conformal Newton-Hooke symmetry. A
possibility to link the equations of motion to a conservation of the
energy-momentum tensor is worth studying as well. The construction
of supersymmetric extensions of the model in this work along the
lines in \cite{JP00,Gal24} is an interesting avenue to explore.

\section*{Acknowledgements}
This work was supported by
the Russian Science Foundation, grant No 23-11-00002.

\end{document}